\documentclass[letterpaper]{article}
\usepackage[margin=1in]{geometry}% Recommended, but optional, packages for figures and better typesetting:
\usepackage{microtype}
\usepackage{graphicx}
\usepackage{subfigure}
\usepackage{booktabs} % for professional tables

\usepackage[utf8]{inputenc} % allow utf-8 input
\usepackage[T1]{fontenc}    % use 8-bit T1 fonts

\usepackage{hyperref}       % hyperlinks
\usepackage{url}            % simple URL typesetting
\usepackage{booktabs}       % professional-quality tables
\usepackage{amsfonts}       % blackboard math symbols
\usepackage{nicefrac}       % compact symbols for 1/2, etc.
\usepackage{microtype}      % microtypography
\usepackage{hyperref}

\usepackage{mathtools}
\usepackage{graphicx,color} %
\usepackage{subfigure} 
\usepackage{algorithm}
\usepackage{amsthm}
\usepackage{amsmath}
\usepackage{bbm}
\usepackage{algorithmic}
\usepackage{hyperref}
\usepackage{amsfonts}
\usepackage{amsmath}
\usepackage[normalem]{ulem}
\usepackage{multicol}
\usepackage{mathtools}
\usepackage{natbib}

\newtheorem{theorem}{Theorem}
\newtheorem{lemma}[theorem]{Lemma}

\newtheorem{definition}{Definition}
\newcommand{\pisim}{\Pi^{\text{sim}}_B}
\newcommand{\piav}{\Pi^{\text{av}}_B}

\DeclarePairedDelimiter\floor{\lfloor}{\rfloor}
\DeclareMathOperator{\sgn}{sgn}

\def\mbs{\boldsymbol}
\def\mc{\mathcal}

\DeclareMathOperator*{\argmin}{argmin}

\newcommand{\bbeta}{\boldsymbol{\beta}}            %
\newcommand{\beps}{\boldsymbol{\epsilon}}            %
\newcommand{\be}{\boldsymbol{e}}            %
\newcommand{\E}{\mathbb{E}}                      %

\newcommand{\bu}{\mathbf{u}}

\newcommand{\Y}{\mathbf{Y}}

\newcommand{\x}{ {\bf x}}
\newcommand{\y}{ {\bf y}}

\newcommand{\tZ}{ {\bf \tilde{Z}}}
\newcommand{\Z}{ {\bf Z}}

\newcommand{\B}{\mathbf{B}}
\newcommand{\C}{\mathbf{C}}

\newcommand{\A}{\mathbf{A}}

\newcommand{\z}{\mathbf{z}}

\renewcommand{\Re}{\mathbb{R}}

\newcommand{\removed}[1]{}
\newcommand{\norm}[1]{\left\| #1 \right\|}

\newcommand{\trans}{{\top}}

\newcommand{\SGI}[1]{\mbs{\beta}_t}

\begin{document}




\title{On Selecting Stable Predictors in Time Series Models}
\author{Avleen S. Bijral}
\maketitle

\begin{abstract}

We propose a novel time series predictor selection scheme that accommodates statistical dependence 
in a more typical i.i.d sub-sampling based selection framework. Furthermore, the machinery of mixing stationary processes allows us to quantify the improved error control of our approach over any base predictor selection method (such as lasso) even in a finite sample setting. Using the lasso as a base procedure we demonstrate the applicability of our methods to simulated and several real time series data. \removed{Finally in the asymptotic regime and when the true data generating process (DGP) is linear, we provide a partial theoretical analysis of the explicit gains achieved by our method over the lasso.}
\end{abstract}

\section{Introduction}

Variable or Predictor selection is an important problem in machine learning and statistics and has received a lot of attention in the literature. These methods are valuable in various applications in biology and finance for both the very large $n$ (samples) or the high dimensional ($p>>n$) setting.  In many cases of interest the objective of variable selection is to uncover an underlying causal mechanism and hence the method in question must exhibit robustness to noise in the data.

Most predictor selection methods discussed in the literature hold only for i.i.d data. The topic is relatively unexplored in the case of time series problems where variables correspond to the choice of exogenous predictors or even the endogenous lags to include in the regression model\footnote{We use predictors to mean either a (possibly lagged) exogenous predictor or a endogenous lag.}.  As time series data becomes pervasive with the advent of sensor enabled devices and networks of these devices (for e.g. \cite{yu2017smart}), such predictor selection is poised to become widely applicable. Some recent applications already employ external time series of web searches to predict disease prevalence such as Influenza (GFT of  \citet{ginsberg2009detecting}, ARGO of \citet{yang2015accurate}) and twitter based sentiment time series for stock prediction \cite{si2013exploiting}. These approaches attempt to mitigate the problem of searching through millions of candidate exogenous time series that may be predictive of the underlying metric.  In such applications selecting the correct time series is particularly crucial as with incorrect predictors, disease or other forecasts can deviate significantly from reality (see \cite{lazer2014parable} for a discussion). \removed{Though \citet{yang2015accurate} attempt to remedy this empirically by using appropriate regularization on the regression parameters of the query time series, no guarantees are provided or even possible. More precisely, the selected predictors must be stable or robust to noise in the data and appear in the selected set for a given method (on multiple subsamples) more often than not.}Another important application lies in the domain of causal inference (\cite{brodersen2015inferring}), wherein the impact of an intervention on a metric is evaluated using a Bayesian structural time series model. This is achieved by forecasting, using metric history and exogenous predictors (pseudo-controls), beyond the time of intervention. Finally, comparing the difference between the forecasts and what really happened, the event's impact is estimated. It is evident that the quality of the forecast is crucial in this task and that the choice of the pseudo-controls or the correct lags is very important and so we would like to have sufficient confidence in this choice.

\subsection{Contribution}

In this paper we propose and analyze novel and efficient stable procedures for predictor selection in time series inspired by the framework developed in \cite{shah2013variable}. To that end we first describe block sampling for time series and then propose stability measures called block pair average (BPA) and simultaneous block selection (SBS) with corresponding error control bounds. We empirically validate our procedures on several real time series and show both qualitative and quantitative improvements over competing methods. \removed{Finally, in case the DGP is known, we provide a partial analysis showing improvement over the lasso base procedure in the asymptotic regime using selection consistency results of \cite{zhao2006model}.} To the best of our knowledge these are the first predictor selection methods with finite sample guarantees that have applications to interpretable forecasting, classification and other time series domains.

\subsection{Related Work}

The foundation of our work is based on extensions to sub-sampling based methods for i.i.d data. In this setting \citet{meinshausen2010stability} proposed Stability Selection (SS) as a repeated sub-sampling based methodology to improve the performance of any variable selection technique and also provided bounds on the number of selected false positives. This method is an improvement over standard variable selection techniques as it is usually non-trivial to provide error control bounds for methods run on the complete data. More recently \citet{shah2013variable} proposed Complementary Pairs Stability Selection (CPSS) procedures that provide significant improvements over SS. Their performance bounds do not explicitly depend on signal and noise variables (that are usually unknown) but instead depends on the number of low selection probability variables that are included and on the number of high selection probability variables that are excluded by the CPSS procedures. This approach doesn't require any dependence on restrictive and unverifiable assumptions such as exchangeability which is required for the analysis in \cite{meinshausen2010stability}. The central idea of stability selection using both SS and CPSS procedures is repeated execution of a base procedure (e.g. lasso - \citet{tibshirani1996regression} ) on subsamples of $\lfloor n/2 \rfloor$ data points to identify variables that show up often in the selected set. In time series applications, the error control yielded by these stability procedures does not hold as the sub-sampling does not account for the underlying dependence.

Most existing predictor selection methods in time series are largely based on heuristics, \cite{ng2013variable} or simply use plain lasso \cite{yang2015accurate,buncic2017macroeconomic} on the entire data and it is non-trivial to provide guarantees for such methods. For the specific case of vector auto regression (VAR) models, \citet{song2011large} propose a grouped penalty based approach that provably identifies relevant lags and predictors in the asymptotic $d$ (number of time series) and $\tilde{p}$ (number of lags) regime. Our method is of a fundamentally distinct flavor in that we provide quantifiable improvement over any base predictor selection method, including the method in \cite{song2011large},  even in the finite data (sample or dimension) setting. Moreover our approach also works for the more general VAR-X (VAR with exogenous) model and in general is independent of the base predictor selection mechanism.

\section{Preliminaries}

In this section we establish notation and preliminary details for the models we work with. Let $\y_t \in \Re^d$ and $\x_t \in \Re^m$ denote strictly stationary sequences. Before we present an example of variable selection procedure in the time series domain first consider the general VAR-X model \cite{Ltkepohl:2007}
\begin{align}
\y_t = \sum_{i=1}^{\tilde{p}} \A_i\y_{t-i} + \sum_{j=0}^s \C_j\x_{t-j} + \mathbf{u}_t
\label{eq:VARXModel}
\end{align}
where each $\A_i \in \Re^{d \times d}$ and $\C_j \in \Re^{d \times m}$. Model \eqref{eq:VARXModel} captures the effect of endogenous and exogenous variables at different lags on the response variable and has been a staple in econometrics with more recent applications in genetics \cite{larvie2016stable} and renewable energy forecasting \cite{cavalcante2016sparse}.

This model can be $\ell_1$-regularized and thus employed as a variable selection procedure, for example to determine stable cross sectional relations in $\y_t$ or to identify predictive exogenous series in $\x_t$. 
\begin{align}
\min_{\B \in d \times (k\tilde{d}+ms)} \sum_{t=\max(\tilde{p},s)}^T L(\y_{t+1}, B\Z_t) + \lambda \norm{\B}_1
\label{eq:VARX}
\end{align}
where $L$ is a convex loss and each $\Z_t \in \Re^{d\tilde{p}+ms}$ and $\B \in \Re^{d \times (dp+ms)}$ are stacked such that 

\begin{align}
&\Z_t = \begin{bmatrix}
\y_t \ldots \y_{t-\tilde{p}+1} \ \x_t  \ldots  \x_{t-s+1}
\end{bmatrix}^{\trans} \notag \\
&\B = \begin{bmatrix}
\A_1 \ldots \A_{\tilde{p}} \ \C_1 \ldots \C_s
\end{bmatrix}
\end{align}

Since $\y_t$ and $\x_t$ is a stationary mixing sequence it follows that $\Z_t$ is also stationary (See \cite{bradley2005}). Unless stated we are not assuming Model \ref{eq:VARX} as the underlying data generating process (DGP). The model and its sparse estimation only serves as a means for predictor selection as described in the later sections.

\section{Stable Predictor Selection}

Consider the stationary sequence $\Z_1,\ldots,\Z_T$ in $\Re^p$ . In the case of  Model \ref{eq:VARX}, $p=d\tilde{p} + ms$. Let us define a variable selection procedure $\hat{S}_{l_T}=\hat{S}_{l_T}(\Z_{i_1},\ldots,\Z_{i_{l_T}})$, as an estimator of the set of signal variables $S \subset \{1,...,p\}$, that takes as input a dependent random sample of length $l_T$ \footnote{Corresponds to the proportion of the entire time series length ($T$) used in sampling. Specified later.} and takes values in all subsets of $\{1,\ldots,p\}$. Define the selection probability of a variable index $k \in \{1,\ldots,p\}$ as
\begin{align}
p_{k,l_T} = \mc{P}(k \in \hat{S}_{l_T}) = \E\left[1_{k \in \hat{S}_{l_T}}\right]
\end{align} 

The improved stability framework as presented in \cite{shah2013variable} executes the base selection procedure $\hat{S}_{\lfloor n/2\rfloor}$ on $B$ i.i.d random samples of size $\lfloor n/2\rfloor$ to then combine the variable selection estimates. In our setting because the stationary sequence is dependent standard stability sub-sampling doesn't apply, so instead using the independent block technique from \cite{yu1994} we create ``almost" independent blocks and then transfer the stability error control to i.i.d blocks that have the same distribution as the original blocks. 

Divide the sequence  $\Z_1,\ldots,\Z_T$ into $2\mu_T$ blocks of length $a_T$ and assume that $T=2\mu_T a_T$ without any loss of generality.  Let $O$ and $E$ be the sets that denote the indices in the odd and even blocks respectively such that 
\begin{align}
&\mathbf{O} = \cup_{j=1}^{\mu_T} O_j \text{,} \notag \\ 
&O_j =\{i : 2(j-1)a_T + 1 \leq i \leq (2j-1)a_T \} \notag \\
&\mathbf{E} = \cup_{j=1}^{\mu_T} E_j \text{,} \notag \\ 
&E_j =\{i : 2(j-1)a_T + 1 \leq i \leq 2ja_T \}
\end{align}
 \begin{figure}%[tbhp]
 \centering
 \includegraphics[scale=0.5]{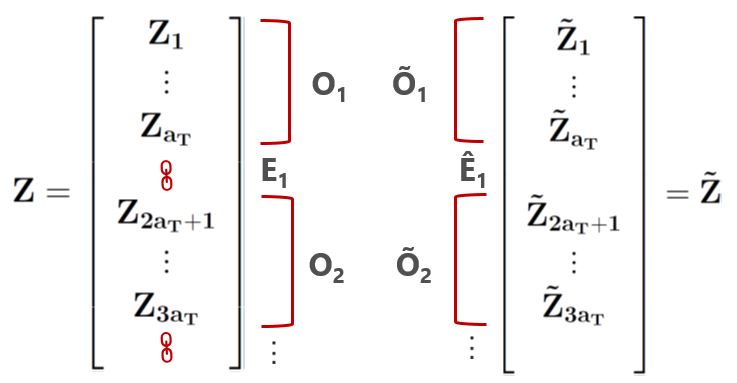}
 \label{fig:blocks}
 \caption{We construct a blockwise independent sequence (right) from a dependent sequence (left) such that the points in the block are dependent but the blocks are independent. The odd blocks in both the sequences have the same distribution.}
 \end{figure}

The intuition behind this approach is that for an appropriately mixing sequence the odd blocks are roughly independent, provided $a_T$ is large enough (See Figure \ref{fig:blocks}). And as we show later in the analysis, if we create ``independent" blocks with each new block having the same distribution as its dependent counterpart we can use this duality to work with the original dependent time series. 

The block construction described leads to measures of stability for predictors that we define next.

\subsection{Definitions}

\subsubsection{Block Pair Average (BPA)}

Consider a randomly selected (without replacement) subset of the sequence of odd blocks  $\mathbf{O}_j = \left(O_{j_1}, \ldots, O_{j_{ \floor{\mu_T/2}}}\right)$.

\begin{definition}{Block Pair Average:}
Let $\left\{ \left(\mathbf{O}_{2j-1},\mathbf{O}_{2j}\right) : j=1,\ldots\, B\right\}$ be the set of randomly selected pairs of sequence of blocks such that $\mathbf{O}_{2j-1} \cap \mathbf{O}_{2j} = \emptyset$. For some $\phi \in  (0,1]$ the block average selection estimator is then defined as $\hat{S}^{\text{av}}=\{ k : \piav(k) \geq \phi \}$ where
\begin{align}
\piav(k) = \frac{1}{2B} \sum_{j=1}^{2B} \mathbbm{1}_{k \in \hat{S}_{\lvert\mathbf{O}_{j}\rvert}} 
\label{eq:simAv}
\end{align}
\end{definition}

Algorithm \ref{alg:main_alg} shows the use of the BPA measure in the context of lasso base predictor.

\subsubsection{Simultaneous Block Selection (SBS)}
\begin{definition}{Simultaneous Block Selection:}
Let $\left\{ \left(\mathbf{O}_{2j-1},\mathbf{O}_{2j}\right) : j=1,\ldots\, B\right\}$ be the set of randomly selected pairs of sequence of blocks such that $\mathbf{O}_{2j-1} \cap \mathbf{O}_{2j} = \emptyset$. For some $\phi \in  (0,1]$ the simultaneous selection estimator is then defined as $\hat{S}^{\text{sim}}=\{ k : \pisim(k) \geq \phi \}$ where
\begin{align}
\pisim(k) = \frac{1}{B} \sum_{j=1}^B \mathbbm{1}_{k \in \hat{S}_{\lvert\mathbf{O}_{2j-1}\rvert}} \mathbbm{1}_{k \in \hat{S}_{\lvert\mathbf{O}_{2j}\rvert}}
\label{eq:simSelec}
\end{align}
\end{definition}

Note that measures \eqref{eq:simAv} and \eqref{eq:simSelec} are dependent data analogues of the i.i.d measures presented in \citet{shah2013variable} and serve the same purpose of reducing variance (via averaging) as estimators of $p_{k,l_T} $. For these we have $l_T=\floor{\mu_T/2}a_T$.

Finally to analyze these measures we need the following definitions
\begin{definition}
For some $\theta \in (0,1]$, define $L_{\theta} = \{k:p_{k,l_T} \leq \theta\}$ be the set of low selection probability features under $\hat{S}_{l_T}$ and $H_{\theta} = \{k:p_{k,l_T} > \theta\}$ denote the set of features that have high selection probability and $\hat{N}_{l_T}=\{1,\ldots,p\}-\hat{S}_{l_T}$. 
\label{def:def3}
\end{definition}

The quantities $\E\left[|\hat{S}^{\text{av}} \cap L_{\theta}|\right] $  and $\E\left[|\hat{N}^{\text{av}} \cap H_{\theta}|\right] $ denote the expected number of low (noise) and high (signal) probability predictors that are included and excluded by our procedure. We analyze these measures against a base predictor to quantify the improvements that our procedures yield.

\subsection{Assumptions}

We will refer to the following assumptions when necessary
\begin{enumerate}
\item \textbf{Assumption 1}: Probability of selection under $\hat{S}_{l_T}$ is better than random. Thus $p_{k,l_T} \geq p_0/p$ where $p_0$ is the number of signal variables. 
\item \textbf{Assumption 2}: The process $\Z_1,\ldots,\Z_T$ is algebraically $\beta$-mixing (See \cite{mohri2009stability}) implying there exists some constant $c_{\beta} > 0$ such that $\beta_t \leq c_{\beta}/t^r$ for $r>0$.
\item \textbf{Assumption 3}: The process $\Z_1,\ldots,\Z_T$ is geometrically $\beta$-mixing (See \cite{bradley2005}) implying there exists some constant $c_{\beta} > 0$ such that $\beta_t \leq \exp(-c_{\beta}t)$. 
\item \textbf{Assumption 4}: The distribution of the sequence $\pisim(k)$ is unimodal for each $k \in L_{\theta}$. This assumption is discussed in Section $3.2$ of \cite{shah2013variable} and reflects the empirical observation that for many different DGPs the distribution of $\pisim(k)$ is consistently unimodal and leads to a sharper version of Markov inequality employed in the analysis.
\end{enumerate}

There are several equivalent definitions of $\beta$-mixing in the literature but we refer to Definition-$2.2$ in \cite{yu1994}.

\section{Block Pair Averaging - Analysis}

For the BPA measure we have the following

\begin{theorem}
If $\phi \in \{\frac{1}{2}+\frac{1}{B},\frac{1}{2}+\frac{3}{2B},...,1\}$ and $\theta < \frac{1}{\sqrt{3}}$, then under Assumptions 1, 2 and 4 and when
\begin{align}
\frac{Tc_\beta}{C(\phi,B) }\max\left(\frac {1}{\left(p_0/p\right)^2},\frac{1}{\left(1-p_0/p\right)^2} \right) \leq a^{r+1}_T , \ r>0
\end{align}
where
\begin{align}
&C(\phi,B) =\notag \\
&\left\{
	\begin{array}{ll}
		\frac{1}{2\left(2\phi-1- 1/(2B) \right)} \mbox{ \text{if}} \\ \phi \in \left(\min\left(\frac{1}{2}+\theta^2, \frac{1}{2}+1/(2B)+\frac{3}{4}\theta^2 \right),\frac{3}{4}\right] \\ \\
		\frac{4\left(1-\phi+1/(2B)\right)}{1+1/B} \mbox{ \text{if} } \phi  \in \left(\frac{3}{4}, 1\right]
	\end{array}
\right.
\end{align}
then we have for the block average procedure \eqref{eq:simAv} and with $l_T=\floor{\mu_T/2}, a_T = \floor{T/4}$
\begin{align*}
\E\left[|\hat{S}^{\text{av}} \cap L_{\theta}|\right] &\leq 2\theta C(\phi,B) \E\left[|\hat{S}_{l_T} \cap L_{\theta}|\right] \\
\E\left[|\hat{N}^{\text{av}} \cap H_{\theta}|\right] & \leq 2\left(1-\theta\right) C(\phi,B) \E\left[|\hat{N}_{l_T} \cap H_{\theta}|\right]
\end{align*}
\label{thm:BlockAvThm}
\end{theorem}
\begin{enumerate}
\item \emph{Remark 1-} $C(\phi,B)$ is the constant in the improved Markov inequality that was developed for a finite grid ($\phi$) in \cite{shah2013variable}.
\item \emph{Remark 2}  When $\phi=0.8$ and $\theta=0.2$ then we get  $\E\left[|\hat{S}^{\text{av}} \cap L_{\theta}|\right]  \leq 0.28 \ \E\left[|\hat{S}_{l_T} \cap L_{\theta}|\right]$, showing that the BPA measure selects only at most $28\%$ of the low selection probability predictors selected by the base procedure.
\item \emph{Remark 3-} For geometric mixing sequences (Assumption 3) we require $a_T=\mc{O}(\log T)$ for the Theorem \ref{thm:BlockAvThm} to hold.
%\item \emph{Remark 3 -} We believe 
%\item \emph{Remark 4 -} The base procedure rely only on a block of length $a_T$.  It is not clear to us how to theoretically analyze stability estimates for longer chunks of the tim
\item \emph{Remark 4-} The improved empirical performance obtained using the $r$-concavity assumption also carries over to our case but for simplicity and generality we retain only the unimodal assumption. See \cite{shah2013variable} for more details.
\item \emph{Remark 5} We can quite easily extend the BPA measure to the case of $d$-BPA wherein we sample $d$ blocks without replacement to get more conservative measures. This becomes obvious in the steps leading upto \eqref{eq:probBound2} in the proof of Theorem \ref{thm:BlockAvThm}. However, for simplicity we retain only the discussion for $d=2$. The i.i.d analogue of this approach seems non-trivial.
\item  \emph{Remark 6} The SBS measure is used in the proof of Theorem \ref{thm:BlockAvThm} and may also be used in practice besides BPA and its analysis is almost exactly similar.
\end{enumerate}
\begin{algorithm} % enter the algorithm environment
\caption{BPA Stable Selection} % give the algorithm a caption
\label{alg:main_alg} % and a label for \ref{} commands later in the document
\begin{algorithmic}[1] % enter the algorithmic environment
    \STATE \textbf{Input:} $Y\in \Re^{T}$, $Z \in \Re^{T \times p}$ , $\phi \in (0.5,0.9]$, $q$, $B$, $a_T$, $\mbs{\lambda}=\{\lambda_1,\ldots,\lambda_{100}\}$ a sequence of regularizers.
\STATE \textbf{Initialize:} $\piav(k) =0$, $\forall k \in \{1,\ldots,p\}$
 \STATE  \textbf{q-Estimate:} Solve $\underset{\mathbf{B}}{\min} \norm{\mathbf{Y} - \mathbf{B}\mathbf{Z}}+ \lambda \norm{\mathbf{B}}_1$ and set $\lambda_q \in \mbs{\lambda}$ to be the smallest $\lambda$ that returns $q$ active entries of $\mathbf{B}$.
   \FOR{$n=1$ {\bfseries to} $B$}
    \STATE \textbf{Sample:} Sequence of blocks $\mc{\mathbf{O}}_1=\{\mathbf{O}_{j_1},\ldots,\mathbf{O}_{j_{\mu_T/2}}\}$ from the set of $\mu_T=\frac{T}{2a_T}$ odd blocks ($\mathbf{O}$) without replacement and set $\mc{\mathbf{O}}_2=\mathbf{O} \setminus \mc{\mathbf{O}}_1$.
   \STATE \textbf{Set:}  For $l \in \{1,2\}$, $\hat{S}_{\lvert \mathbf{O}_{l}\rvert} =  \{i :   \hat{\mathbf{B}}_i \neq 0, \hat{\mathbf{B}} = \underset{\mathbf{B}}{\argmin} \norm{\mathbf{Y}(\mc{\mathbf{O}}_l) - \mathbf{B}\mathbf{Z}(\mc{\mathbf{O}}_l) }^2 + \lambda_q \norm{\mathbf{B}}_1\}$.
\STATE \textbf{BPA:}
$\piav(k) = \piav(k)+\mathbbm{1}_{k \in \hat{S}_{\lvert \mathbf{O}_{1}\rvert}}/(2B) + \mathbbm{1}_{k \in \hat{S}_{\lvert\mathbf{O}_{2}\rvert}}/(2B)$, $\forall k \in \{1,\ldots,p\}$.
   \ENDFOR
\STATE \textbf{Output:} $\hat{S}^{\text{av}}=\{ k : \piav(k) \geq \phi \}$
\end{algorithmic}
\end{algorithm}
\section{Empirical Results}

To convince the reader of the utility of our methods we evaluate them on simulated and real time series data using lasso as a base predictor. For the real time series we evaluate forecast errors on a hold-out test period using the selected predictors from Algorithm \ref{alg:main_alg} and compare them to other predictor selection methods such as standard lasso, elastic net \cite{zou2005regularization}, adaptive lasso \cite{zou2006adaptive} and AIC based maximum lag selection for both $\tilde{p}$ and $s$ (See Model \ref{eq:VARXModel}). For these methods we use the standard method of cross-validation. Note that the standard method of CV works for auto-regressive models as long as the errors are assumed to be uncorrelated. See \cite{bergmeir2018note} for a discussion.

The post selection training is restricted to least squares regression. The train-test split is $67\%/33\%$ and the predictions are made in a rolling fashion with each test data point added to the training set and re-trained to forecast the next step. On the simulated data we test the robustness of our method by adding varying degree of noise to the simulated data and reporting true positive and false positive rates (TPR/FPR). We de-trend the real data in case of any obvious violations of stationarity.

Before we get into the details of the empirical results we first discuss the choice of parameters and how to use Theorem \ref{thm:BlockAvThm} for the case of lasso as a base selection method.

\subsection{Theory in Practice}\label{subsec:T_I_P}
Since $\E\left[|\hat{S}_{l_T} \cap L_{\theta}|\right] $ is not known in practice we approximate it by $q=\E\left[|\hat{S}_{l_T}|\right]$. Setting $q$ is equivalent to choosing the regularization parameter $\lambda$, we can vary it until lasso selects $q$ predictors. \removed{Just like in the i.i.d version of stable selection (See \cite{meinshausen2010stability} and \cite{shah2013variable}), it is our experience that our results are not that sensitive to the choice of $q$ (or $\lambda$).}Once $q$ is set we use $\theta=q/p$ to denote the irrelevant variables as those having less than average selection probability.

The appropriate threshold can then be determined based on Theorem \ref{thm:BlockAvThm} by solving for $\phi$ 
\begin{align}
2\theta C(\phi,B) \E\left[|\hat{S}_{l_T} \cap L_{\theta}|\right] &\leq 2\theta qC(\phi,B) \notag \\
& = \frac{2q^2C(\phi,B)}{p} \notag =  lp
\end{align}
where $l \in (0,1]$ is pre-specified.  In general we can specify any two of $q$, $\phi$ and $l$ to get the stable predictor set. We will present results with different values of $q$ for some threshold $\phi \in (0.5,0.9]$. $B$ is set to $50$ and $a_T$ was set to some multiple of seasonality in the data for all experiments. This reflects the empirical assumption that across seasons the data points have a weak dependence. In the absence of seasonality we didn't see a significant difference against other choices of $a_T$ such as $\sqrt{T}$ or $\log T$. See Algorithm \ref{alg:main_alg} for a complete recipe of our method using the lasso base predictor. In the algorithm $\mathbf{Y}(\mathbf{O}_l)$ and $\mathbf{Z}(\mathbf{O}_l)$ correspond to the data points from the corresponding block sequence (see also the proof of Theorem \ref{thm:BlockAvThm}). The regularizer sequence is the commonly used in the glmnet package \cite{glmnet}. Note that in the proofs we require $\lfloor T/4 \rfloor$ of the data to be used for the comparison against the base procedure, however in Algorithm \ref{alg:main_alg} and empirically we use all the data for determining the $q$-estimated active set.

 \begin{figure}%[tbhp]
 \centering
 \includegraphics[width=.49\linewidth]{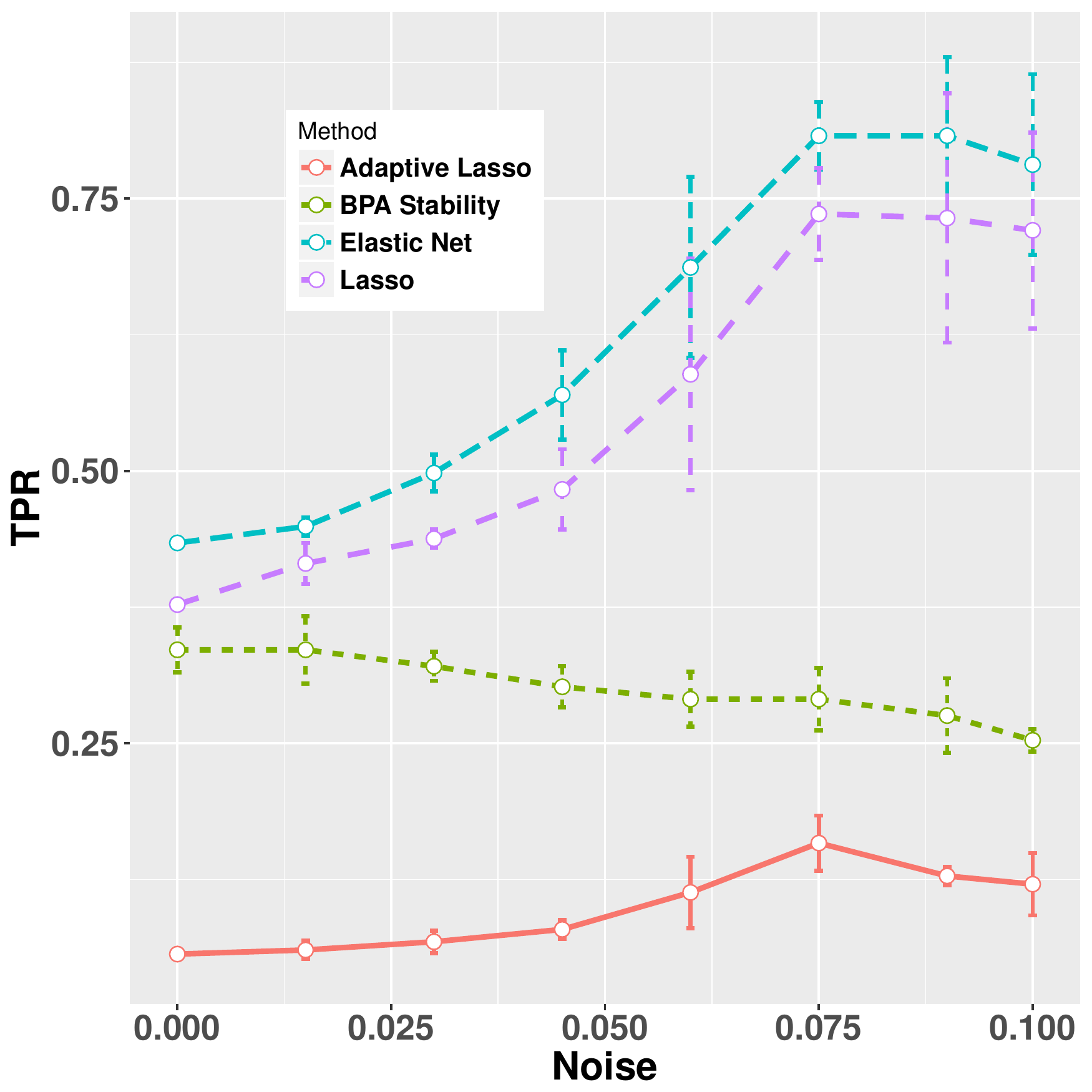}
 \includegraphics[width=.49\linewidth]{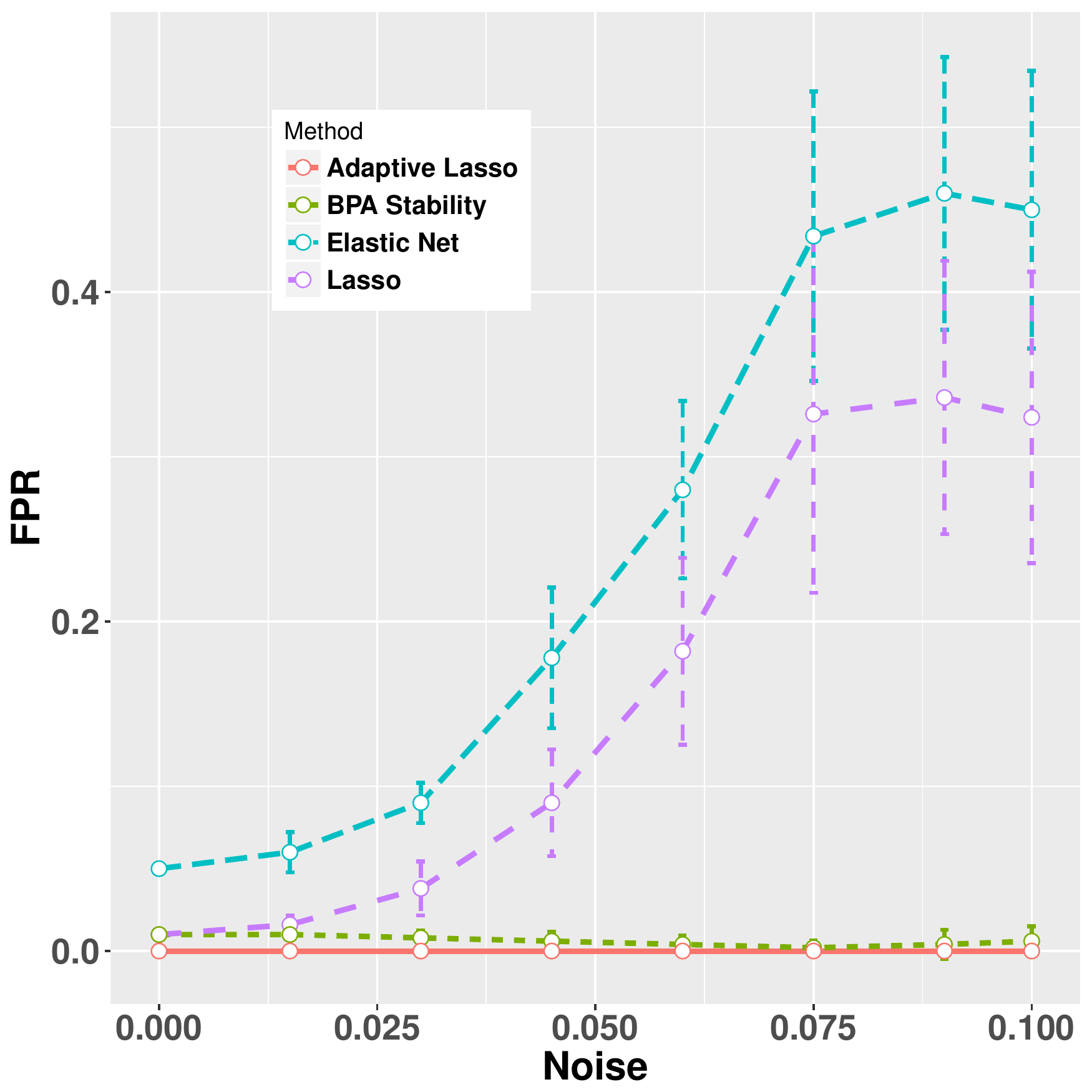}
  \includegraphics[scale=0.4]{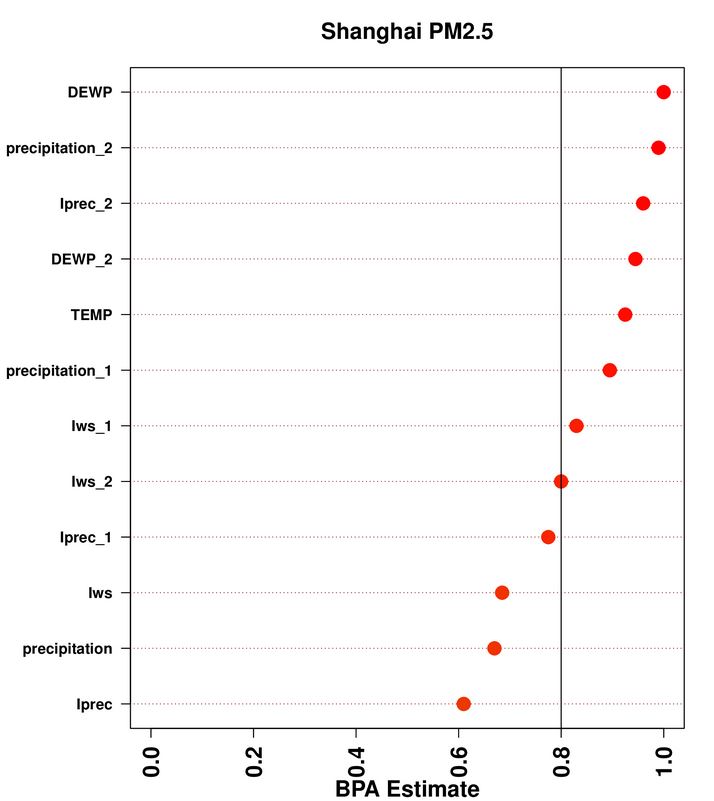}
 \caption{a) The TPR/FPR reveals that Algorithm \ref{alg:main_alg} is much less sensitive to added noise and hence more robust. The competing procedures are either too conservative (adaptive lasso) or too accepting. b) BPA ranked predictors. Dew point is the top predictor and there is a causal mechanismm for this. Defined as the temperature at which dew droplets start to form and so not surprisingly this impacts the PM2.5 measurement significantly.}
  \label{fig:TPR_FPR_BPA}
 \end{figure}
 \begin{table}[t]
\centering
\begin{tabular}{|l|l|l|l|l|l|}
\hline
\textbf{Method} &\textbf{BeijingPM2.5} &\textbf{ShanghaiPM2.5} & \textbf{IndoorTemp} & \textbf{AirQuality}  \\ \hline
$q_1$-BPA           &$1.946/294.159$ & $\mathbf{2.052/44.844}$& $\mathbf{1.197/4.345}$       & $\mathbf{1.875/4.487}$            \\ \hline
$q_1$-Lasso        & $1.994/\mathbf{292.555}$  & $2.267/49.415$&  $1.209/4.385$       & $2.247/5.608$            \\ \hline
$q_2$-BPA           &$2.368/333.700$ & $2.616/56.98$& $1.324/4.837$       & $3.744/9.621$            \\ \hline
$q_2$-Lasso        & $2.424/343.126$  & $2.873/62.607$&  $1.446/5.403$       & $4.116/10.455$            \\ \hline
Lasso           & $2.584/359.337$ & $3.490/77.106$& $1.507/5.657$       & $5.728/15.086$           \\ \hline
ElasticNet         & $2.561/357.331$ & $3.487/76.937$&  $1.508/5.657$       & $5.716/15.01$        \\ \hline
AdapLasso          &$2.597/360.919$ & $3.497/77.139$& $1.212/4.396$       & $5.527/14.366$         \\ \hline
AIC         &$\mathbf{1.868}/300.788$& $3.209/70.978$& $1.271/4.739$       & $5.665/14.804$          \\ \hline
\end{tabular}
\caption{RMSE/MAPE on the test period compared to competing methods. Here $q_1=\lfloor0.2 p \rfloor$ and $q_2=\lfloor 0.6 p \rfloor$.}
\label{tab:rmse-data}
\end{table}
\subsection{Simulated Data}

We simulate a synthetic time series with $T=10000$ from the following auto-regressive with exogenous variables (AR(3)-X) model where the exogenous time series are simulated from a simple AR($2$) model with Gaussian innovations.
\begin{align}
Y_t = \sum_{i=1}^3 \theta_i Y_{t-i} + \sum_{j=1}^{50} \beta_j X^j_{t} + \epsilon_t \sim \mc{N}(0,\sigma^2)
\label{eq:sim_model}
\end{align}

We also simulate $100$ time series (AR($3$)) for the dataset along with $2$ i.i.d noise time series to create a more realistic setting of several candidates with varying possibility of being included in the stable set. Our objective is to recover a stable set (lags and exogenous) using Algorithm \ref{alg:main_alg} for a AR(3)-X model with $104$ exogenous time series. The parameters are normally distributed, $\beta_i \sim \mc{N}(0,0.05)$ for all $i=1,...,50$.

We vary the amount of noise $(\mc{N}(0,\sigma^2))$ added to the predictors and the response and plot the TPR/FPR with $1$-standard deviation error bars. To estimate the stable predictors using the BPA measure we set $q=40$ and $\phi=0.8$. Figure \ref{fig:TPR_FPR_BPA}a) shows that the stable set is more robust to growing noise levels compared to the competing approaches. The TPR for lasso and Elastic net is higher since it selects many more predictors and consequently also has a much higher FPR. In contrast adaptive lasso is too conservative and Algorithm \ref{alg:main_alg} has an FPR comparable to it but much better TPR. Note that for $\sigma^2=0$ only less than half of the original predictors in Model \eqref{eq:sim_model} show up in the selected set, since a pure least squares estimation of the model reveals that only a fraction of the $50$ predictors are statistically significant.

\section{Real Data}

\subsection{Chinese Cities PM2.5}

The Chinese cities PM2.5 dataset, \cite{PM2_5} is an hourly sampled measurement of particulate matter concentration alongside several other exogenous meteorological time series. Our goal is to forecast the PM2.5 concentration of Beijing and Shanghai using exogenous and endogenous signals. The data specific parameters settings are $T=52854$, $a_T=3*24$ (thrice the daily seasonality), $\tilde{p}_{\text{max}}=24$ and $s_{\text{max}}=3$ (max endogenous and exogenous lags to select from).

\subsection{Indoor Temperature}

The indoor temperature dataset \cite{SML2010} is a one minute sampled ($15$ minute smoothed) set of measurements by a monitoring system that captures various attributes such as CO2 concentration, humidity, lighting, outdoor conditions etc. The idea behind this dataset is to aid in accurate forecasting of the indoor temperature  to better regulate the energy consumption of the HVAC system. For this analysis we only consider the impact of the exogenous signal on the temperature so that $\tilde{p}_{\text{max}}=0$ and $s_{\text{max}}=2$. Other parameters are $T=4137$, $a_T=96$. 

\subsection{Air Quality - Temperature Prediction}

The air quality dataset \cite{airq} is an hourly sampled set of measurements by chemical sensors that capture various attributes such as CO, NO2 and O3 concentration. While the original purpose of this dataset is to estimate sensor estimation quality, we repurpose it to model temperature as a function of these concentrations. For this analysis we set $\tilde{p}_{\text{max}}=24$ and $s_{\text{max}}=3$. Other parameters are $T=9358$, $a_T=24$. 

\subsection{Results and Discussion}

Table \ref{tab:rmse-data} reveals that the BPA measure performs better on all but one datasets. For the BeijingPM2.5 dataset it appears that all variables contribute to the improved RMSE. Note that the choice of $q$ does make a difference to the results, as too small a value would exclude signal variables (appears to be the case for BeijingPM2.5) and a large $q$ would add noisy variables to the selected set. Some prior knowledge of the domain can be useful here, but in its absence a largish $q$ (for example $\lfloor 0.5p \rfloor$) seems to be a safe choice (due to the guaranteed error control for the BPA measure) as the results indicate that for most datasets RMSE only degrades a little.
From a causality perspective Figure \ref{fig:TPR_FPR_BPA}b) reveals that dew point (temperature at which dew droplets form) and precipitation are important variables. This is corroborated by meteorological studies such as \cite{liang2015assessing}. 

\section{Conclusion}

Filling the gap in the feature selection literature for dependent data, we proposed novel stable predictor selection techniques in the time series setting. For robustness scores that can be used with any selection method, we provided theoretical results guaranteeing error control. The stable predictors selected by our method were shown to have superior predictive performance on several real datasets. In the future we intend to explore such a scheme and measures for nonstationary and heteroscedastic time series.

\section{Appendix}

We gather all the proofs in this section 

\subsection{Proof of Theorem \ref{thm:BlockAvThm}}
\begin{proof}
Denote the sequence of random variables that correspond to the $\mathbf{O}_j$ and $\mathbf{E}_j$ indices as
\begin{align}
\Z(\mathbf{O}_j) = \{\Z_i, i \in \mathbf{O}_j\}, \ \Z(\mathbf{E}_j) = \{\Z_i, i \in \mathbf{E}_j\}
\end{align}
 
Consider a sequence of i.i.d blocks $\{\tZ(\mathbf{O}_j): j=1,\ldots,\mu_T\}$ where $\tZ(\mathbf{O}_j)=\{\tZ_i: i \in \mathbf{O}_j\}$ such that the sequence is independent of $\Z_1,\ldots,\Z_T$ and each block has the same distribution as a block from the original sequence ($\Z(\mathbf{O}_j)$). 

Let $h(\mathbf{O})$ be a bounded measurable function on the set of selected blocks and $\tilde{\mathbf{O}}$ be the corresponding i.i.d sequence of blocks from $\tZ(\mathbf{O}_j)$ then
\begin{align}
\E\left[h(\mathbf{O})\right] &= \E\left[h(\mathbf{O})\right] - \E\left[h(\tilde{\mathbf{O}})\right] + \E\left[h(\tilde{\mathbf{O}})\right] \notag \\ 
&\leq \E\left[\left|h(\mathbf{O}) - h(\tilde{\mathbf{O}})\right|\right] + \E\left[h(\tilde{\mathbf{O}})\right] 
\end{align}
where the expectation is w.r.t to the distribution of the original and constructed sequences.
For a measurable and bounded function $h$ such that  $|h| \leq M$ we have from Lemma 4.1 of \cite{yu1994}
\begin{align}
\E\left[h(\mathbf{O})\right] &\leq M(\mu_T-1)\beta_{a_T} + \E\left[h(\tilde{\mathbf{O}})\right] 
\label{eq:IBBnd}
\end{align}
Next we have
\begin{align}
0 &\leq \frac{1}{B} \sum_{j=1}^B \left(1-\mathbbm{1}_{k \in \hat{S}_{\lvert\mathbf{O}_{j_1}\rvert}}\right )\left(1-\mathbbm{1}_{k \in \hat{S}_{\lvert\mathbf{O}_{j_2}\rvert}}\right) \notag \\
&= 1-2\piav(k)+\pisim(k)
\label{eq:sims_bound}
\end{align}

Using \eqref{eq:IBBnd} for $h(\mathbf{O})=\mathbbm{1}_{\pisim(k) \geq 2\phi-1}$, using the unimodal Markov inequality for the simultaneous selector on the constructed independent blocks $\{\tilde{\mathbf{O}}\}$, from Theorem 3 of \citep{shah2013variable} and finally using the fact that the sequence is stationary (all blocks have the same distribution)
\begin{align}
\mc{P}(k \in \hat{S}^{\text{av}})&=\mc{P}\left(\piav(k) \ge \phi \right) \notag \\
&\leq \mc{P}\left(\pisim(k) \geq 2\phi-1\right) \notag \\
&=  \E\left[\mathbbm{1}_{\pisim(k) \geq 2\phi-1}\right] \notag \\
&\leq (\mu_T-1)\beta_{a_T} + \E\left[h(\tilde{\mathbf{O}})\right] \notag \\
&= (\mu_T-1)\beta_{a_T} + \mc{P}_{\tilde{O}}\left(\pisim(k) \geq 2\phi-1\right) \notag \\
&\leq (\mu_T-1)\beta_{a_T} + C(\phi,B)\E_{\tilde{\mc{O}}}\left[\pisim(k)\right] \notag \\
&= (\mu_T-1)\beta_{a_T} + C(\phi,B)p^2_{k,l_T}
\label{eq:probBound2}
\end{align}
where 
\begin{align}
&C(\phi,B) =\notag \\
&\left\{
	\begin{array}{ll}
		\frac{1}{2\left(2\phi-1- 1/(2B) \right)} \mbox{ \text{if}} \\ \phi \in \left(\min\left(\frac{1}{2}+\theta^2, \frac{1}{2}+1/(2B)+\frac{3}{4}\theta^2 \right),\frac{3}{4}\right] \\ \\
		\frac{4\left(1-\phi+1/(2B)\right)}{1+1/B} \mbox{ \text{if} } \phi  \in \left(\frac{3}{4}, 1\right]
	\end{array}
\right.
\end{align}

Using the assumption $\left(p_0/p\right) \leq p_{k,a_T}$ (Assumption 1) it is easy to see that when $\mu_T(c_\beta/a^r_T) \leq C(\phi,B) p^2_{k,l_T}$
\begin{align}
%\frac{\max\left(\log T, 2\log \frac{p}{p_0}\right)}{c_{\beta}} \leq a_T 
\frac{Tc_\beta p^2}{2C(\phi,B) p^2_0} \leq a^{r+1}_T 
\label{eq:Tbnd2}
\end{align}
This implies from \eqref{eq:probBound2} that when \eqref{eq:Tbnd2} holds
\begin{align}
\mc{P}(k \in \hat{S}^{\text{av}}) \leq 2 C(\phi,B) p^2_{k,l_T}
\label{eq:probBound3} 
\end{align}
Following the proof of Theorem $1$ in \cite{shah2013variable} and using \eqref{eq:probBound3} we have
\begin{align}
\E\left[|\hat{S}^{\text{av}} \cap L_{\theta}|\right] &= \E\left [ \sum_{k=1}^p \mathbbm{1}_{k \in \hat{S}^{\text{av}}} \mathbbm{1}_{p_{k,l_T} \leq \theta} \right] \notag \\
& = \sum_{k=1}^p \mc{P}(k \in \hat{S}^{\text{av}}) \mathbbm{1}_{p_{k,l_T} \leq \theta} \notag \\
& \leq 2C(\phi,B)\sum_{k=1}^p p^2_{k,l_T} \mathbbm{1}_{p_{k,l_T} \leq \theta} \notag \\
& \leq 2C(\phi,B)\theta \left(\sum_{k=1}^p p_{k,l_T} \mathbbm{1}_{p_{k,l_T} \leq \theta} \right) \notag \\
& = 2\theta C(\phi,B) \E\left[|\hat{S}_{l_T} \cap L_{\theta}|\right]
\end{align}
For the second part replace $\hat{S}^\text{av}$ by $\hat{N}^{\text{av}}$ as an estimator of $N$, the set of noise variables. Let $\Pi^{\text{av}}_{B,\hat{N}}$ and $\Pi^{\text{sim}}_{B,\hat{N}}$ correspond to the noise variable estimators. Let $h(\tilde{\mathbf{O}})=\mathbbm{1}_{\Pi^{\text{sim}}_{B,\hat{N}}}$
\begin{align}
\mc{P}(k \in \hat{N}^{\text{av}}) &=\mc{P}(k \notin \hat{S}^{\text{av}}) \notag \\
& = \mc{P}\left(\piav(k) < \phi \right)\notag \\ 
&= \mc{P}\left(\Pi^{\text{av}}_{B,\hat{N}} > 1-\phi \right) \notag \\
&\leq\mc{P}\left(\Pi^{\text{sim}}_{B,\hat{N}} \geq 2\phi-1\right) \notag \\
&=  \E\left[\mathbbm{1}_{\Pi^{\text{sim}}_{B,\hat{N}} \geq 2\phi-1}\right] \notag \\
&\leq (\mu_T-1)\beta_{a_T} + \E\left[h(\tilde{\mathbf{O}})\right] \notag \\
&= (\mu_T-1)\beta_{a_T} + \mc{P}_{\tilde{O}}\left(\Pi^{\text{sim}}_{B,\hat{N}} \geq 2\phi-1\right) \notag \\
&\leq (\mu_T-1)\beta_{a_T} + C(\phi,B)\E_{\tilde{\mc{O}}}\left[\Pi^{\text{sim}}_{B,\hat{N}}\right] \notag \\
&= (\mu_T-1)\beta_{a_T} + C(\phi,B)\left(1-p_{k,l_T}\right)^2
\label{eq:probBound_N}
\end{align}
Just like in the proof of the first part when $Tc_\beta /\left(2C(\phi,B) \left(1-p_0/p\right)^2\right) \leq a^{r+1}_T $
\begin{align}
\mc{P}(k \in \hat{N}^{\text{av}}) \leq 2 C(\phi,B) \left(1-p_{k,l_T}\right)^2
\label{eq:probBound3_N} 
\end{align}
\begin{align}
&\E\left[|\hat{N}^{\text{av}} \cap H_{\theta}|\right] \notag \\
&= \E\left [ \sum_{k=1}^p \mathbbm{1}_{k \in \hat{N}^{\text{av}}} \mathbbm{1}_{p_{k,l_T} > \theta} \right]\notag \\
&= \sum_{k=1}^p \mc{P}(k \in \hat{N}^{\text{av}}) \mathbbm{1}_{p_{k,l_T} > \theta} \notag \\
&\leq 2C(\phi,B)\sum_{k=1}^p \left(1-p_{k,l_T}\right)^2 \mathbbm{1}_{p_{k,l_T} > \theta} \notag \\
& \leq 2\left(1-\theta\right)C(\phi,B)\left(\sum_{k=1}^p \left(1-p_{k,l_T}\right) \mathbbm{1}_{p_{k,l_T} > \theta} \right) \notag \\
& = 2\left(1-\theta\right) C(\phi,B) \E\left[|\hat{N}_{l_T} \cap H_{\theta}|\right]
\end{align}

\end{proof}
\newpage
\bibliographystyle{plainnat}
\bibliography{opt}

\begin{thebibliography}{26}
\providecommand{\natexlab}[1]{#1}
\providecommand{\url}[1]{\texttt{#1}}
\expandafter\ifx\csname urlstyle\endcsname\relax
  \providecommand{\doi}[1]{doi: #1}\else
  \providecommand{\doi}{doi: \begingroup \urlstyle{rm}\Url}\fi

\bibitem[PM2()]{PM2_5}
Chinese pm2.5.
\newblock
  \url{https://archive.ics.uci.edu/ml/datasets/PM2.5+Data+of+Five+Chinese+Cities}.

\bibitem[SML()]{SML2010}
Smart house data.
\newblock \url{https://archive.ics.uci.edu/ml/datasets/SML2010}.

\bibitem[air()]{airq}
Air quality.
\newblock \url{http://archive.ics.uci.edu/ml/datasets/Air+quality}.

\bibitem[Bergmeir et~al.(2018)Bergmeir, Hyndman, and Koo]{bergmeir2018note}
Christoph Bergmeir, Rob~J Hyndman, and Bonsoo Koo.
\newblock A note on the validity of cross-validation for evaluating
  autoregressive time series prediction.
\newblock \emph{Computational Statistics \& Data Analysis}, 120:\penalty0
  70--83, 2018.

\bibitem[Bradley(2005)]{bradley2005}
Richard~C. Bradley.
\newblock Basic properties of strong mixing conditions. a survey and some open
  questions.
\newblock \emph{Probab. Surveys}, 2:\penalty0 107--144, 2005.
\newblock \doi{10.1214/154957805100000104}.
\newblock URL \url{http://dx.doi.org/10.1214/154957805100000104}.

\bibitem[Brodersen et~al.(2015)Brodersen, Gallusser, Koehler, Remy, Scott,
  et~al.]{brodersen2015inferring}
Kay~H Brodersen, Fabian Gallusser, Jim Koehler, Nicolas Remy, Steven~L Scott,
  et~al.
\newblock Inferring causal impact using bayesian structural time-series models.
\newblock \emph{The Annals of Applied Statistics}, 9\penalty0 (1):\penalty0
  247--274, 2015.

\bibitem[Buncic and Tischhauser(2017)]{buncic2017macroeconomic}
Daniel Buncic and Martin Tischhauser.
\newblock Macroeconomic factors and equity premium predictability.
\newblock \emph{International Review of Economics \& Finance}, 2017.

\bibitem[Cavalcante et~al.(2016)Cavalcante, Bessa, Reis, and
  Dowell]{cavalcante2016sparse}
Laura Cavalcante, Ricardo~J Bessa, Marisa Reis, and Jethro Dowell.
\newblock Sparse structures for very short-term wind power forecasting.
\newblock \emph{working paper}, 2016.

\bibitem[Friedman et~al.(2010)Friedman, Hastie, and Tibshirani]{glmnet}
Jerome Friedman, Trevor Hastie, and Robert Tibshirani.
\newblock Regularization paths for generalized linear models via coordinate
  descent.
\newblock \emph{Journal of Statistical Software}, 33\penalty0 (1):\penalty0
  1--22, 2010.
\newblock URL \url{http://www.jstatsoft.org/v33/i01/}.

\bibitem[Ginsberg et~al.(2009)Ginsberg, Mohebbi, Patel, Brammer, Smolinski, and
  Brilliant]{ginsberg2009detecting}
Jeremy Ginsberg, Matthew~H Mohebbi, Rajan~S Patel, Lynnette Brammer, Mark~S
  Smolinski, and Larry Brilliant.
\newblock Detecting influenza epidemics using search engine query data.
\newblock \emph{Nature}, 457\penalty0 (7232):\penalty0 1012--1014, 2009.

\bibitem[Larvie et~al.(2016)Larvie, Sefidmazgi, Homaifar, Harrison,
  Karimoddini, and Guiseppi-Elie]{larvie2016stable}
Joy~Edward Larvie, Mohammad~Gorji Sefidmazgi, Abdollah Homaifar, Scott~H
  Harrison, Ali Karimoddini, and Anthony Guiseppi-Elie.
\newblock Stable gene regulatory network modeling from steady-state data.
\newblock \emph{Bioengineering}, 3\penalty0 (2):\penalty0 12, 2016.

\bibitem[Lazer et~al.(2014)Lazer, Kennedy, King, and
  Vespignani]{lazer2014parable}
David Lazer, Ryan Kennedy, Gary King, and Alessandro Vespignani.
\newblock The parable of google flu: traps in big data analysis.
\newblock \emph{Science}, 343\penalty0 (6176):\penalty0 1203--1205, 2014.

\bibitem[Liang et~al.(2015)Liang, Zou, Guo, Li, Zhang, Zhang, Huang, and
  Chen]{liang2015assessing}
Xuan Liang, Tao Zou, Bin Guo, Shuo Li, Haozhe Zhang, Shuyi Zhang, Hui Huang,
  and Song~Xi Chen.
\newblock Assessing beijing's pm2. 5 pollution: severity, weather impact, apec
  and winter heating.
\newblock \emph{Proceedings of the Royal Society A: Mathematical, Physical and
  Engineering Sciences}, 471\penalty0 (2182):\penalty0 20150257, 2015.

\bibitem[Ltkepohl(2007)]{Ltkepohl:2007}
Helmut Ltkepohl.
\newblock \emph{New Introduction to Multiple Time Series Analysis}.
\newblock Springer Publishing Company, Incorporated, 2007.
\newblock ISBN 3540262393, 9783540262398.

\bibitem[Meinshausen and B{\"u}hlmann(2010)]{meinshausen2010stability}
Nicolai Meinshausen and Peter B{\"u}hlmann.
\newblock Stability selection.
\newblock \emph{Journal of the Royal Statistical Society: Series B (Statistical
  Methodology)}, 72\penalty0 (4):\penalty0 417--473, 2010.

\bibitem[Mohri and Rostamizadeh(2009)]{mohri2009stability}
Mehryar Mohri and Afshin Rostamizadeh.
\newblock Stability bounds for stationary $\phi$-mixing and $\beta$-mixing
  processes.
\newblock \emph{Journal of Machine Learning Research}, 4:\penalty0 1--26, 2009.

\bibitem[Ng et~al.(2013)]{ng2013variable}
Serena Ng et~al.
\newblock Variable selection in predictive regressions.
\newblock \emph{Handbook of economic forecasting}, 2\penalty0 (Part
  B):\penalty0 752--789, 2013.

\bibitem[Shah and Samworth(2013)]{shah2013variable}
Rajen~D Shah and Richard~J Samworth.
\newblock Variable selection with error control: another look at stability
  selection.
\newblock \emph{Journal of the Royal Statistical Society: Series B (Statistical
  Methodology)}, 75\penalty0 (1):\penalty0 55--80, 2013.

\bibitem[Si et~al.(2013)Si, Mukherjee, Liu, Li, Li, and Deng]{si2013exploiting}
Jianfeng Si, Arjun Mukherjee, Bing Liu, Qing Li, Huayi Li, and Xiaotie Deng.
\newblock Exploiting topic based twitter sentiment for stock prediction.
\newblock \emph{ACL (2)}, 2013:\penalty0 24--29, 2013.

\bibitem[Song and Bickel(2011)]{song2011large}
Song Song and Peter~J Bickel.
\newblock Large vector auto regressions.
\newblock \emph{arXiv preprint arXiv:1106.3915}, 2011.

\bibitem[Tibshirani(1996)]{tibshirani1996regression}
Robert Tibshirani.
\newblock Regression shrinkage and selection via the lasso.
\newblock \emph{Journal of the Royal Statistical Society. Series B
  (Methodological)}, pages 267--288, 1996.

\bibitem[Yang et~al.(2015)Yang, Santillana, and Kou]{yang2015accurate}
Shihao Yang, Mauricio Santillana, and Samuel~C Kou.
\newblock Accurate estimation of influenza epidemics using google search data
  via argo.
\newblock \emph{Proceedings of the National Academy of Sciences}, 112\penalty0
  (47):\penalty0 14473--14478, 2015.

\bibitem[Yu(1994)]{yu1994}
Bin Yu.
\newblock Rates of convergence for empirical processes of stationary mixing
  sequences.
\newblock \emph{Ann. Probab.}, 22\penalty0 (1):\penalty0 94--116, 01 1994.
\newblock \doi{10.1214/aop/1176988849}.
\newblock URL \url{http://dx.doi.org/10.1214/aop/1176988849}.

\bibitem[Yu et~al.(2017)Yu, Mak, Zhang, Wong, Zheng, Lau, and
  Poon]{yu2017smart}
Ruoxi Yu, Tony~WC Mak, Ruikai Zhang, Sunny~H Wong, Yali Zheng, James~YW Lau,
  and Carmen~CY Poon.
\newblock Smart healthcare: Cloud-enabled body sensor networks.
\newblock In \emph{Wearable and Implantable Body Sensor Networks (BSN), 2017
  IEEE 14th International Conference on}, pages 99--102. IEEE, 2017.

\bibitem[Zou(2006)]{zou2006adaptive}
Hui Zou.
\newblock The adaptive lasso and its oracle properties.
\newblock \emph{Journal of the American statistical association}, 101\penalty0
  (476):\penalty0 1418--1429, 2006.

\bibitem[Zou and Hastie(2005)]{zou2005regularization}
Hui Zou and Trevor Hastie.
\newblock Regularization and variable selection via the elastic net.
\newblock \emph{Journal of the royal statistical society: series B (statistical
  methodology)}, 67\penalty0 (2):\penalty0 301--320, 2005.

\end{thebibliography}

\end{document}